# Laser Annealing of Transparent ZnO Thin Films: A Route to Improve Electrical Conductivity and Oxygen Sensing Capabilities

A. Frechilla[1,2], J. Frechilla[1], L. A. Angurel[1]*, F. Toldrá-Reig[2], E. Martínez[1], G. F. de La Fuente[1] and D. Muñoz-Rojas[2]*

[1] Instituto de Nanociencia y Materiales de Aragón, CSIC-Universidad de Zaragoza, María de Luna 3, E-50018 Zaragoza, Spain;
[2] Univ. Grenoble Alpes, CNRS, Grenoble INP, LMGP, F-38000 Grenoble, France;

**\*Corresponding authors**
Luis Alberto Angurel Lambán: angurel@unizar.es
David Muñoz-Rojas: david.munoz-rojas@grenoble-inp.fr

**ABSTRACT**
The chemical deposition of high-performance Zinc Oxide (ZnO) thin films is challenging, thus significant efforts have been devoted during the past decades to develop cost-effective, scalable fabrication methods in gas phase. This work demonstrates how ultra-short-pulse Laser Beam Scanning (LBS) can be used to modulate electrical conductivity in ZnO thin films deposited on soda-lime glass by Spatial Atomic Layer Deposition (SALD), a high-throughput, low-temperature deposition technique suitable for large-area applications. By systematically optimizing laser parameters, including pulse energy and hatching distance, significant improvements in the electrical performance of 90 nm-thick ZnO films were achieved. The optimization of the laser annealing parameters —0.21 µJ/pulse energy and a 1 µm hatching distance— yielded ZnO films with an electrical resistivity of $(9 \pm 2) \cdot 10^{-2}$ Ω·cm, 3 orders of magnitude lower than as deposited films. This result suggests that laser post-deposition-processing can play an important role in tailoring the properties of ZnO thin films. Excessive laser intensity can compromise structural integrity of the films, however, degrading their electrical transport properties.
Notably, the electrical resistance of laser-annealed ZnO films exhibited high sensitivity to oxygen concentration in the surrounding atmosphere, suggesting exciting prospects for application in devices based on transparent oxygen sensors. This study thus positions ultra-short pulsed laser annealing as a versatile post-deposition method for fine-tuning the properties of ZnO thin films, enabling their use in advanced optoelectronic and gas-sensing technologies, particularly on temperature-sensitive substrates.

**Keywords**: Zinc oxide (ZnO), Spatial Atomic Layer Deposition (SALD), laser annealing, oxygen sensing

## GRAPHICAL ABSTRACT

(1) Laser irradiation of a ZnO thin film 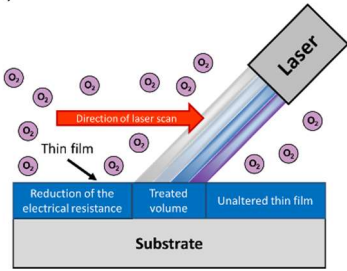

(2) Enhanced electrical behaviour 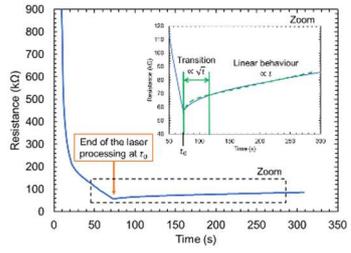

(3) Oxygen sensing capabilities 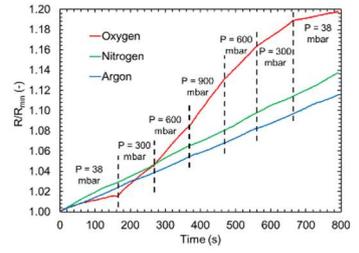

# 1.- Introduction

Zinc oxide (ZnO) is a versatile semiconductor material that has attracted significant interest in both fundamental research and applied science due to its exceptional physical and chemical properties [1]. With a wide direct bandgap of approximately 3.6 eV [2] and a high exciton binding energy of 60 meV [3,4], ZnO exhibits promising optoelectronic and photonic performance, making it suitable for a variety of applications.

Its inherent transparency, non-toxicity, chemical stability, low cost and high durability further contribute to the potential of transparent conductive oxides (TCOs) [5] in applications which include gas sensing [6–9], photocatalysis [10–12], spintronics [13] and short-wavelength optoelectronic devices [14,15]. Additionally, ZnO is being increasingly explored in next-generation energy technologies, including hybrid perovskite solar cells, where it plays a key role in enhancing long-term device stability and power conversion efficiency [16–18]. These trends underscore the growing interest in ZnO across numerous technological fields. In many of these applications, the use of flexible substrates is essential for future industrial adoption. It is thus crucial to develop growth and processing methods that operate at low temperatures, ensuring compatibility with flexible, polymer-based substrates.

Various deposition techniques have been employed to prepare ZnO thin films, including solution approaches [19], magnetron sputtering [20,21], chemical vapour deposition (CVD) [3] and pulsed laser deposition (PLD) [22,23]. However, significant challenges still remain in coating large areas with high-quality thin films, which poses a key hurdle for industrial-scale production. Many efforts have also been made to develop cost-effective fabrication methods for ZnO thin films, but those produced through low-cost thermal processes often demonstrate limited properties [24]. In this context, Spatial Atomic Layer Deposition (SALD) has gathered attention due to its capability to deposit high-quality nanometre-thick materials over large surfaces with faster growth rates compared to conventional ALD [25,26]. Moreover, SALD can also be performed at atmospheric pressure, offering the same fine thickness control and high conformality typical of ALD, thus making it more suitable for scaling up to meet industrial and commercial demands [27,28]. As a result, SALD emerges as a highly promising deposition technique for high-quality ZnO thin films.

An effective method for improving thin film performance at low temperatures involves post-treatment of ZnO thin films [5]. Post-annealing processes frequently enhance the physical properties of materials, especially in TCOs [29,30]. High-temperature annealing can eliminate defects and promote recrystallization, thereby improving the quality and properties of the films. However, conventional resistance heating techniques often require extended operation times, consume substantial energy and are incompatible with polymeric or other soft substrates [31].

Laser annealing offers a convenient solution to this problem [32,33]. By focusing photon energy on the surface of the ZnO thin film, the laser can selectively anneal the thin film while minimizing heat exposure to the surrounding areas [34]. With optimal laser parameters, the desired modifications can be induced in the thin film without damaging the substrate [35–39]. Furthermore, laser annealing can be applied over large areas with precise control over the irradiated regions, enabling patterning of selected areas under a variety of processing conditions. In addition, appropriate laser beam scanning results in recrystallization of the ZnO film [40], as recently demonstrated for other functional metallic oxides [41]. A comprehensive study is thus convenient to assess how different laser processing parameters affect the electrical properties of ZnO and essential towards enhancing its physical and chemical properties. In particular, the use of ultraviolet (UV) wavelengths may significantly improve the electrical properties of thin films, as illumination with a UV radiation has suggested that adsorbed oxygen species at grain boundaries may be neutralized [42], releasing oxygen molecules [5] and creating oxygen vacancies within the ZnO lattice. Simultaneously, it may enhance the gas sensing response by promoting electron–hole pair generation, increasing oxygen molecule adsorption, and extracting more electrons from the conduction band compared to non-UV conditions [8,43–45]. This is because the 3.6 eV ZnO band gap closely matches the energy of available UV laser wavelengths (i.e. $\lambda$ = 355 nm).

This work proposes an innovative approach, centred on low-temperature, high-throughput growth (SALD) and UV laser post-processing annealing, to produce high-quality ZnO thin films using scalable production techniques. The influence of relevant laser annealing parameters, in particular, laser energy per pulse and pulse overlap, on the electrical properties of 90 nm-thick ZnO films deposited on soda-lime glass substrates is studied. The temporal evolution of the electrical resistivity after laser processing in different atmospheres is also assessed. Notably, the irradiated films show a high and reversible sensitivity to oxygen, a property we leveraged to successfully demonstrate their functionality as transparent, room-temperature oxygen sensors with a detection limit of 30 mbar constrained by our measurement set-up.

## 2.- Experimental details

**Film Preparation.** ZnO thin films were deposited on commercial soda-lime glass substrates with a homemade SALD system. Additional information about the SALD technique and the experimental setup can be also found in other sources [26,28,46]. Diethylzinc (($C_2H_5$)$_2$Zn; DEZ, Aldrich), and water vapour were used as precursors for zinc and oxygen, respectively. The ZnO deposition process involved 600 cycles for estimated film thickness of ca. 90 ± 5 nm, based on the growth per cycle (GPC) of the process used at 65 °C, while the precursors were kept at room temperature. The substrate oscillated under the injector at a distance of 150 µm with a speed of 10 cm/s. The film thickness was also estimated using a FS-1 Multi-Wavelength compact ellipsometer. The obtained films were cut in 1 x 1 cm$^2$ squares (*W*) for this study.

**Post-deposition Treatment.** The annealing processes applied to the SALD deposited samples were performed using an ultraviolet pulsed laser (Rofin-Sinar, Germany) with an emission wavelength $\lambda$ = 355 nm, and a pulse duration $\tau$ = 300 ps. The laser provides a maximum output power of 2.3 W with a linearly polarized beam output. The beam exhibits an elliptical Gaussian profile with 1/e$^2$ intensity decay dimensions [47] of 2$a$ = 34 µm and $b/a$ = 0.86 at the working distance.

The laser is equipped with galvanometric mirrors so that the spot can define a 1-dimensional scan of length $l_L$ along the surface of the sample at different beam speeds, $v_L$. Unless otherwise indicated, a speed $v_L$ = 800 mm/s, and a pulse repetition frequency $f_{rep}$ = 800 kHz, were applied in this work (i.e. the distance between two consecutive pulses in one line was 1 µm). In order to scan a 2-dimensional (2D) region, the system allows to define the distance between consecutive beam scanning lines, i.e. the hatching distance, $\delta$. Laser processing was performed in beam scanning mode, without moving the sample. Laser pulse energy, $E_p$, as well as pulse and line-to-line overlap were varied in order to improve the properties of the resulting films.

Unless otherwise indicated, the samples were processed in an evacuated chamber at pressures of 30-38 mbar using a rotary pump. Additional experiments were performed with the films under different air, nitrogen and argon pressures.

**Film Characterization.** Thin film electrical resistance ($R$) was measured with a Keithley 2000 multimeter connected to a 4-point probe station. In the case of the laser *operando* measurements, a 2-point method was used, as shown schematically in **figure 1**. In the latter measurement configuration, two homemade electrodes (2 mm-wide copper strips) were carefully positioned to rest fully on the film, each covering an entire side of the sample. The distance between both voltage taps was fixed at $L \approx 4.5$ mm. The resistance of the electrodes in short-circuit was determined as < 2 Ω, which is negligible compared to the resistance of the film. Unless otherwise indicated, the electrical measurements were performed at room temperature in ambient atmosphere.

The surface microstructure was characterised using a FEG ZEISS Gemini 300 field-emission scanning electron microscope (SEM) (Carl Zeiss, Jena, Germany) operated at 10 kV.

## 3. Modelling of the laser annealing process

Laser processing can be studied so as to assess the induced phenomena occurring during *operando* measurements. A schematic representation of the beam scanning mode and the *operando* measurement set-up is illustrated in **figure 1**.

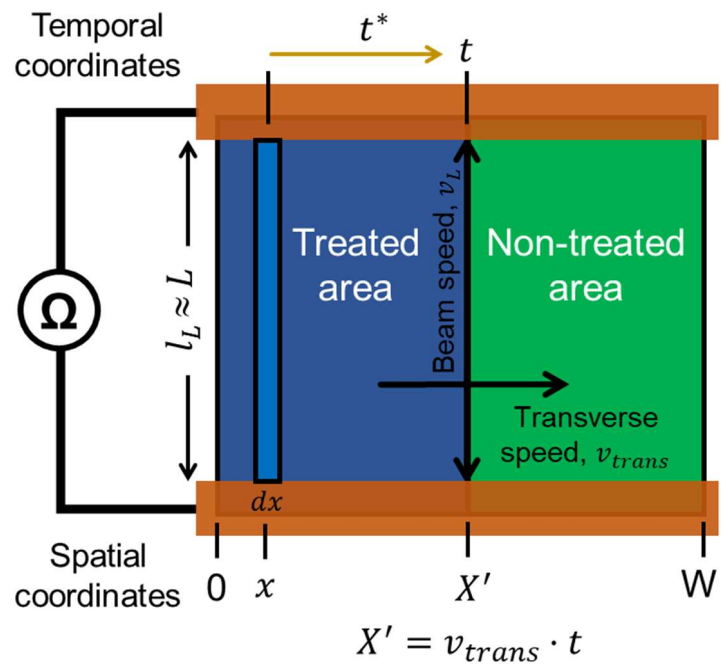

*Figure 1. Scheme of the UV laser beam scan process applied to anneal ZnO thin films, during operando measurement of their resistance with 2-points. The laser beam scans a line of length $l_L$ and the equivalent transverse velocity of the laser front, as explained in the text, is defined by $v_{trans}$. The voltage taps (orange rectangles) are separated a distance L ($\approx l_L$, length of the 1-D scan). W is the film width.*

The 2D laser scanning process performed in this experiment is closely related to the hatching distance ($\delta$), in turn determining an equivalent transverse velocity of the laser front, $v_{trans}$. This velocity is related to the length of a one-dimensional scan ($l_L$) and the beam speed ($v_L$) through the relationship presented in **equation 1**.

$$\frac{l_L}{v_L} = \frac{\delta}{v_{trans}} \qquad (1)$$

The temporal evolution of the complete sample resistance, while the laser is irradiating the film, can be evaluated considering the following. Once the laser irradiates a given area of the sample, the resistance of such area is suddenly lowered during a short initial stage, and it is followed by a slow increase thereafter. With this assumption and considering the measurement set-up, the sample resistance is evaluated as the contribution of two resistances in parallel that vary with time: the first one associated to the untreated region ($R_{untreated}$), and the second one to the laser-treated region ($R_{treated}$), as illustrated in **equation 2**.

$$\frac{1}{R(t)} = \frac{1}{R_{untreated}(t)} + \frac{1}{R_{treated}(t)} \qquad (2)$$

The resistivity of the untreated region ($\rho_0$) is considered to be that of the as-deposited ZnO film. For the calculation of the resistivity of the laser-treated regions ($\rho_1$), the sample was divided into differential regions of width $dx$ that were processed in a given time. As mentioned earlier, each of these regions exhibits a strong initial reduction of the resistivity due to the laser irradiation and evolves with time from the end of the laser treatment until the measurement takes place. By defining $t^*$ as the time between the instant when the laser irradiates a certain $dx$ region and the measuring time ($t$) (see **figure 1**), the following equation can be written (**equation 3**):

$$t^* = t - \frac{x}{v_{trans}}, \qquad (3)$$

The resistivity $\rho_1(t^*)$ would be obtained from **equation 4**.

$$\rho_1(t^*) = \beta(t^*)\rho_0 = \beta_0 f_1(t^*)\rho_0 \text{ with } \beta_0 < 1 \qquad (4)$$

where $\beta(t^*)$ is the function that models the time evolution of each $dx$ portion of the film during and after the laser irradiation. $\beta_0$ accounts for the initial resistivity reduction and $f_1(t^*)$ considers the time evolution of the resistance after the laser irradiation. Therefore, the time evolution of the resistance of the laser treated area can be evaluated by **equation 5**.

$$\frac{1}{R_{treated}(t)} = \int_0^{X'} \frac{e}{\beta_0 \rho_0 L} \frac{dx}{f_1(t^*)}, \qquad (5)$$

where $X'$ is the position of the laser scanning line at time, $t$, $e$ is the film thickness, and $L$ the distance between contacts. On the other hand, the resistance of the untreated region can be directly calculated using the following relation (**equation 6**):

$$\frac{1}{R_{untreated}(t)} = \frac{(W - v_{trans} \cdot t) \cdot e}{\rho_0 \cdot L} \tag{6}$$

Where $W$ refers to the total width of the film.

## 4.- Results and discussion
### 4.1. Evolution of the electrical properties of ZnO films with varying laser processing parameters

**Figure 2** illustrates the evolution of the electrical response of a SALD ZnO thin film as a function of energy per pulse for different hatching distances. The remaining laser processing parameters are outlined in the experimental section. As deduced from this figure, large variations of the electrical resistance of the ZnO films can be achieved by controlling laser annealing. More specifically, the initial insulating behaviour of the as-deposited ZnO films is modified as the pulse energy increases, leading to reduced sheet resistance values. For optimum processing conditions ($E_p$ = 0.21 µJ and $\delta$ = 1 µm) the resistance decreases down to 115 kΩ. Considering the measurement set-up, the lowest electrical resistivity values $\rho$ = (9 ± 2) · $10^{-2}$ Ω cm for the laser annealed ZnO thin films correspond to electrical conductivity values of 11 ± 6 $\Omega^{-1}cm^{-1}$ under optimum conditions.

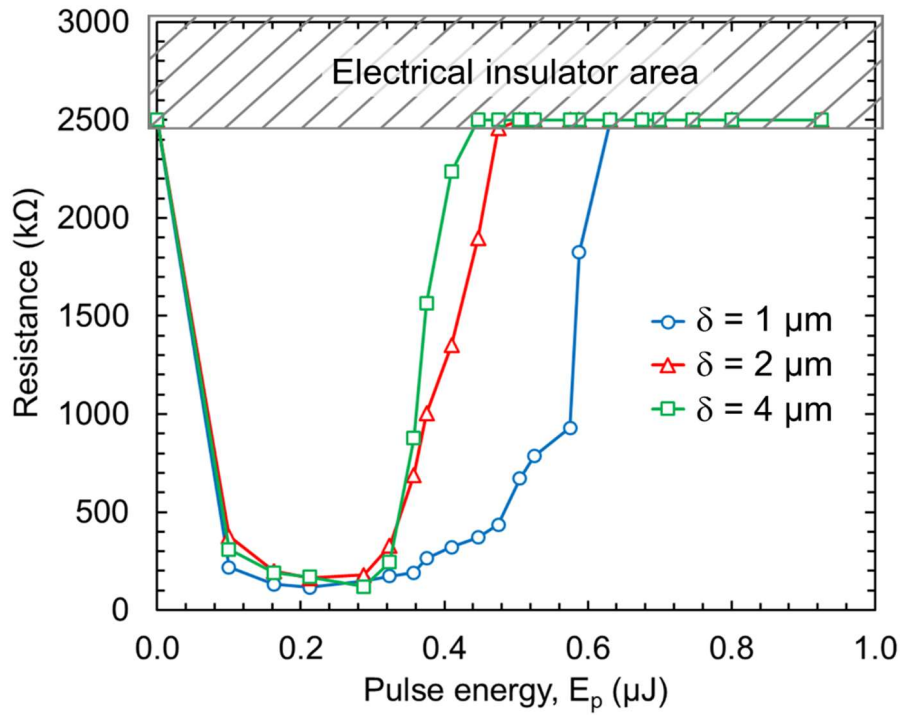

*Figure 2. Resistance measured by 4-point method in ZnO thin films after laser annealed as a function of the pulse energy for different hatching distances, as indicated. Lines are placed as eye-guides.*

As deduced from **Figure 2**, the laser parameters exert a strong influence on the resistance values of the films, although a critical optimum condition may not be identified, i.e. there is a range of pulse energies and hatching distances which yield improved electrical properties for these films. Thus, for $\delta$ = 2 µm and 4 µm, this window ranges from 0.1 to 0.4 µJ/pulse, while this range becomes broader for $\delta$ = 1 µm, extending from 0.1 to 0.6 µJ/pulse. It is thus clear that both, pulse energy and the spacing between consecutive laser lines, significantly influence the evolution of the electrical properties. Hatching determines the 2D pulse overlap, which in turn governs various physical processes during laser annealing. The best performance observed for the smallest hatching distance can be attributed to increased heat accumulation during laser processing upon reducing $\delta$, also enhanced by a pulse duration in the hundreds of picoseconds. At the same pulse energy, a shorter hatching distance results in greater heat accumulation and prolonged high-temperature exposure, leading to slower cooling rates. This effect helps preserve the integrity of the film, thereby extending the range of pulse energies that yield enhanced electrical properties.

**Figure 3** shows the SEM images of the ZnO film surface as-deposited and after UV laser irradiation with $\delta$ = 1 µm and different pulse energy values. A comparison of **figure 3 (a)** and **(b)** confirms that the low laser energy treatment (0.21 µJ) that provided the best electrical results preserves the integrity of the film. In contrast, pulse energies exceeding the optimal range mentioned above result in significant changes in the films. Consequently, the progressive formation of domains on the sample surface, which reduce percolation between ZnO islands (**figure 3 (c)**), can explain the increase of resistance observed for this condition. The surface topography resulting from higher pulse energies (**figure 3 (d)**) suggests the occurrence of melting and recrystallization processes. A large molten volume triggers an increased influence of surface tension in relation with the film-substrate interaction forces. These phenomena enhance the formation of melt drops and a consequent progressive separation of ZnO domains, which result in the ultimate formation of separate ZnO islands, in agreement with the electrical insulating behaviour observed in **figure 2** for this condition.

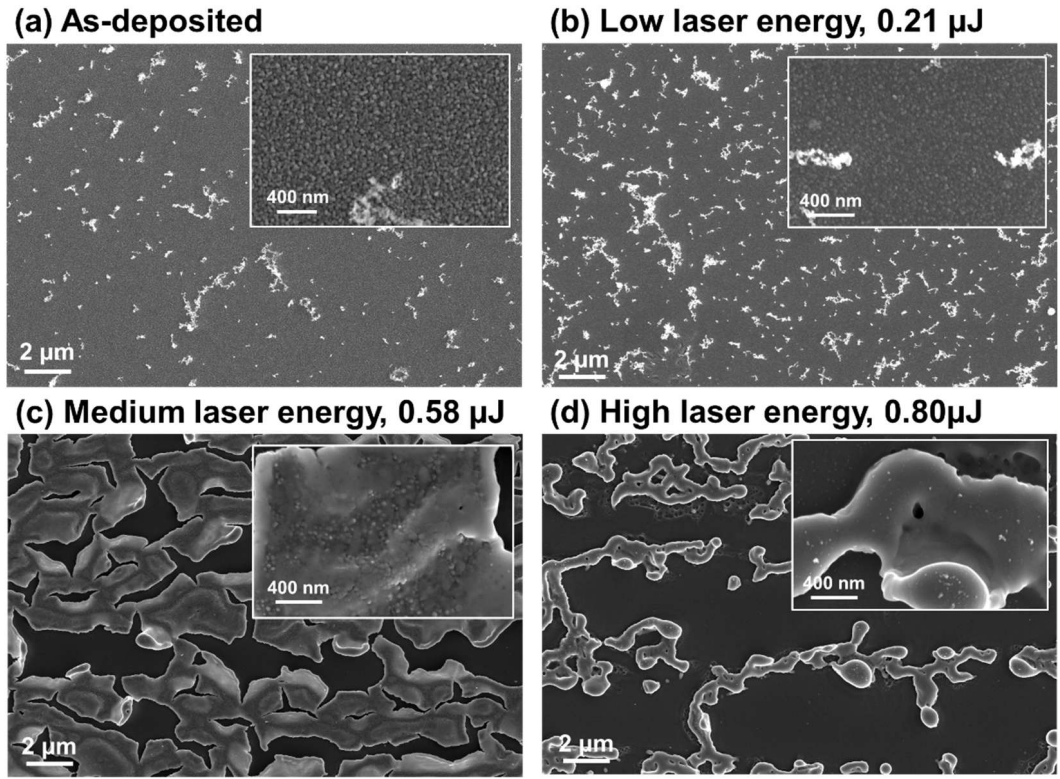

*Figure 3.* SEM micrographs (in lens detector) of representative areas of ZnO-film surface after laser annealing using $\delta$ = 1 µm and different pulse energies: (a) as-deposited, (b) 0.21 µJ, (c) 0.58 µJ and (d) 0.80 µJ. Insets show higher magnification images.

In essence, appropriate UV laser irradiation conditions induce electrical conductivity without producing holes or other type of detectable damage in ZnO SALD films. For low (optimum) laser pulse energies, the thin film appears to undergo other type of modifications.

Most oxides can lose oxygen from their structure under intense irradiation and high local temperatures [48–50], conditions that occur during UV laser irradiation. These processes lead to the generation of oxygen vacancies and the desorption of trapped oxygen from the deposition process [5,49], which are responsible for the observed enhancement in electrical behaviour. Moreover, as it was previously mentioned, since the laser wavelength used (355 nm) closely matches the energy of the ZnO band gap (3.6 eV), it may promote electron–hole pair generation, extracting more electrons from the valence band [8,42–44]. The obtained resistivity values in this study are competitive with those reported in the literature for thicker thin films. For undoped ZnO thin films, different fabrication methods including spray and plasma deposition yield electrical conductivities of 6.74 $\Omega^{-1}cm^{-1}$ for 460 nm films [51], 42 $\Omega^{-1}cm^{-1}$ for 550 nm films [52], and 40 $\Omega^{-1}cm^{-1}$ for 190 nm films [53]. Conductivities of 1.4 $\Omega^{-1}cm^{-1}$ have been reported for 600 nm La:ZnO films [54], while higher values of up to 345 $\Omega^{-1}cm^{-1}$ have been achieved for 185 nm Al:ZnO films [5]. Other high temperature (> 200 ºC) SALD ZnO thin films exhibit conductivity values in the range of 20 – 250 $\Omega^{-1}cm^{-1}$ [55–57]. For example, SALD ZnO thin films with a thickness of 210 nm show a conductivity of 25 $\Omega^{-1}cm^{-1}$ under UV irradiation. However, doping ZnO with aluminium can again decrease resistivity, yielding conductivities as high as 173 $\Omega^{-1}cm^{-1}$ [58].

## 4.2.- Resistance time evolution

Given the improvements in electrical conductivity induced by UV laser processing, a more in-depth study of the changes taking place during the laser irradiation process was performed using *in situ* and *operando* electrical resistance measurements, as explained in section 3.

**Figure 4** shows the temporal evolution of the electrical resistance of a ZnO thin film during and after the laser treatment. The laser processing parameters were: $E_p$ = 0.3 µJ/pulse, $v_L$ = 800 mm/s, $f_{rep}$ = 800 kHz and $\delta$ = 1 µm. The two regions that were described in section 3 can be clearly identified.

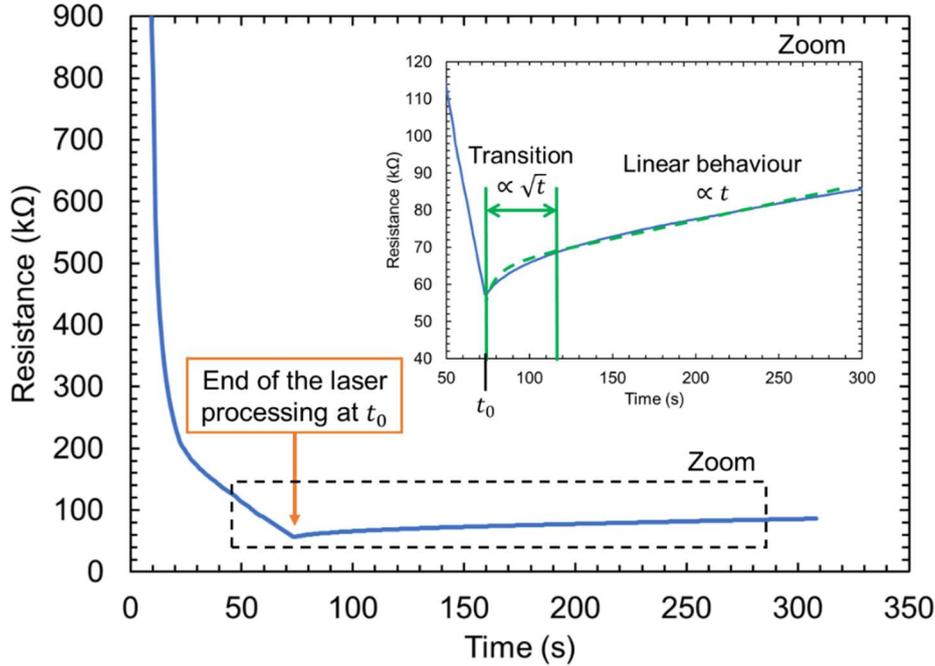

**Figure 4.** *In situ measurement of a ZnO thin film during and after laser annealing ($\delta$ = 1 µm and $E_p$ = 0.3 µJ, $v_L$ = 1000 mm/s, $f_{rep}$ = 800 kHz). The inset shows a zoomed view of the last part of the curve. The dashed lines in the inset correspond to square-root and linear fittings with time, as explained in the text.*

The elevated temperatures reached during the laser treatment, combined with the increased surface reactivity of the sample, result in an important increase in the resistance of ZnO films immediately after the laser treatment. Two regions have been indeed observed in the time evolution of the film resistance after laser annealing, as observed in the inset of **figure 4**. During the initial 45 s, the resistance increases at a faster rate, following closely a $\sqrt{t - t_0}$ dependence, where $t_0$ is the time when the laser treatment is stopped. In a second step, the resistance variation rate stabilizes, leading to a linear variation with rates in the range between 0.1 and 1 kΩ/s. These depend on the laser treatment conditions. This increase in the resistance values is maintained until the material becomes again fully insulating. This trend can be attributed either to the loss of oxygen vacancies [59,60] generated during the laser treatment, or to the subsequent reabsorption of oxygen. Both effects play a crucial role in the sample's electrical properties [5,43,44,49].

The time-dependent variation of the resistance during the *operando* measurements is evaluated with the model presented in **section 3**. Based on the experimental results, the $f_1(t^*)$ dependence introduced in equation 4 can be modelled as: $f_1(t^*) = 1 + \alpha\sqrt{t^*}$ with an $\alpha$ parameter ($\alpha > 0$). Thus, the resistivity $\rho_1(t^*)$ would be given by:

$$\rho_1(t^*) = \beta f_1(t^*)\rho_0 = \beta_0(1 + \alpha\sqrt{t^*})\rho_0 \text{ with } \beta_0 < 1; \alpha > 0 \tag{7}$$

Therefore, the time evolution of the resistance of the laser treated area can be evaluated solving the following **equation**:

$$\frac{1}{R_{treated}(t)} = \int_0^{X'} \frac{e}{\beta_0\rho_0 L} \frac{dx}{\left(1 + \alpha\sqrt{t - \frac{x}{v_{trans}}}\right)} \tag{8}$$

Solving **equation 8** using the procedure described in the Supplementary information, $R_{treated}(t)$ follows this dependence:

$$\frac{1}{R_{treated}(t)} = \frac{-2 \cdot v_{trans}\, e}{\beta_0\rho_0 L}\left(\frac{1}{\alpha}\sqrt{t} - \frac{1}{\alpha^2}\ln(1 + \alpha\sqrt{t})\right) \tag{9}$$

Combining **equations 6** and **9**, the total resistance of the sample at a given time *t* is described in **equation 10**.

$$\frac{1}{R(t)} = \frac{1}{R_{initial}}\left[1 - \frac{v_{trans}}{W}\cdot t + \frac{2\cdot v_{trans}}{\beta_0 W \alpha}\left(\sqrt{t} - \frac{1}{\alpha}\ln(1 + \alpha\sqrt{t})\right)\right] \tag{10}$$

The measured *operando* electrical resistance reduction during UV laser annealing and the predicted calculations from the proposed model are compared in **figure 5**. The curve of the proposed model matches closely the trend observed on the experimental results, with fitting parameters $\beta_0 = 9.7 \cdot 10^{-4}$ and $\alpha = 3.14\ \text{s}^{-1/2}$ for these processing conditions.

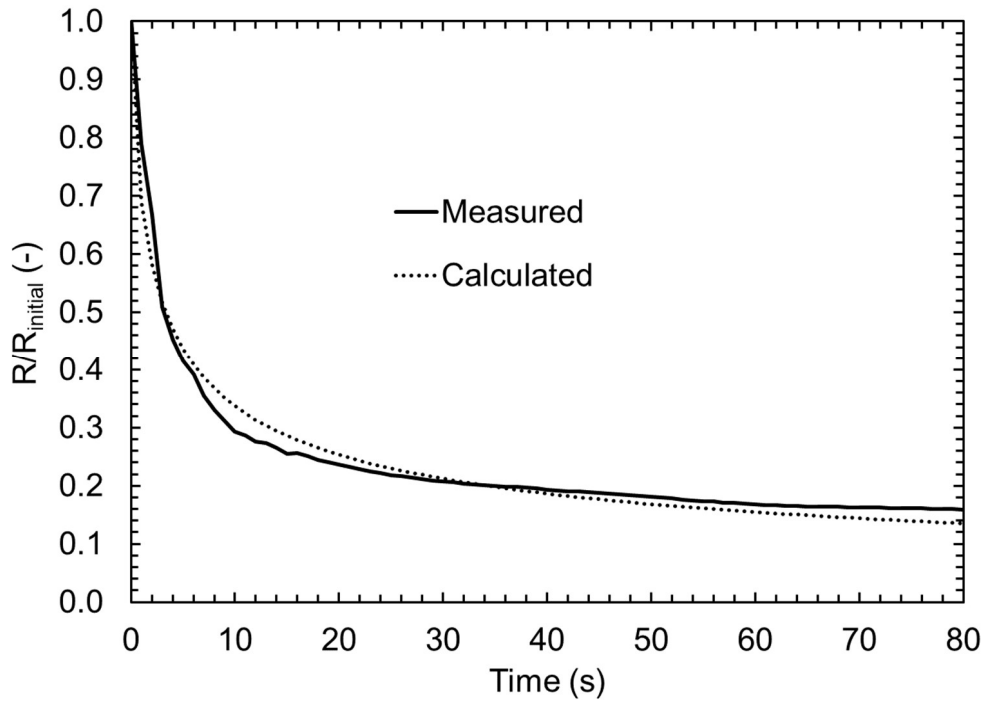

*Figure 5. Comparison of the time dependence of the relative resistance, measured and calculated for operando conditions during laser processing. The electrical measurement was performed while the laser processed area was increasing.*

Overall, these measurements yield another significant outcome by demonstrating the selectivity inherent to laser processing, as it stands out as a prominent advantage of this processing route. Consequently, the changes in the electrical properties induced by the laser are limited exclusively to the irradiated area, with possible adverse effects observed in the heat-affected zone. In contrast, the electrical behaviour of the non-treated area remains unaltered under all circumstances.

Given the evolution of the resistance after the laser treatment, and in order to understand and exploit it, further experiments were performed. First we evaluated the reversibility of the process upon further laser treatments. A set of experiments was performed where the laser was alternatively turned on and off while the resistance ($R$) was measured. In these experiments, the laser was turned on and it irradiated the ZnO until the full thin film was treated, reaching at that stage a minimum value ($R_{min}$). This process took different times for a 1 cm$^2$ sample depending on the laser conditions used, namely, 64, 32 and 16 s for the three hatching distances used of 1 µm, 2 µm and 4 µm, respectively. Afterwards, the laser was turned off during 64 s and the increase in the resistance was evaluated. This process was repeated several times and it is shown in **figure 6**.

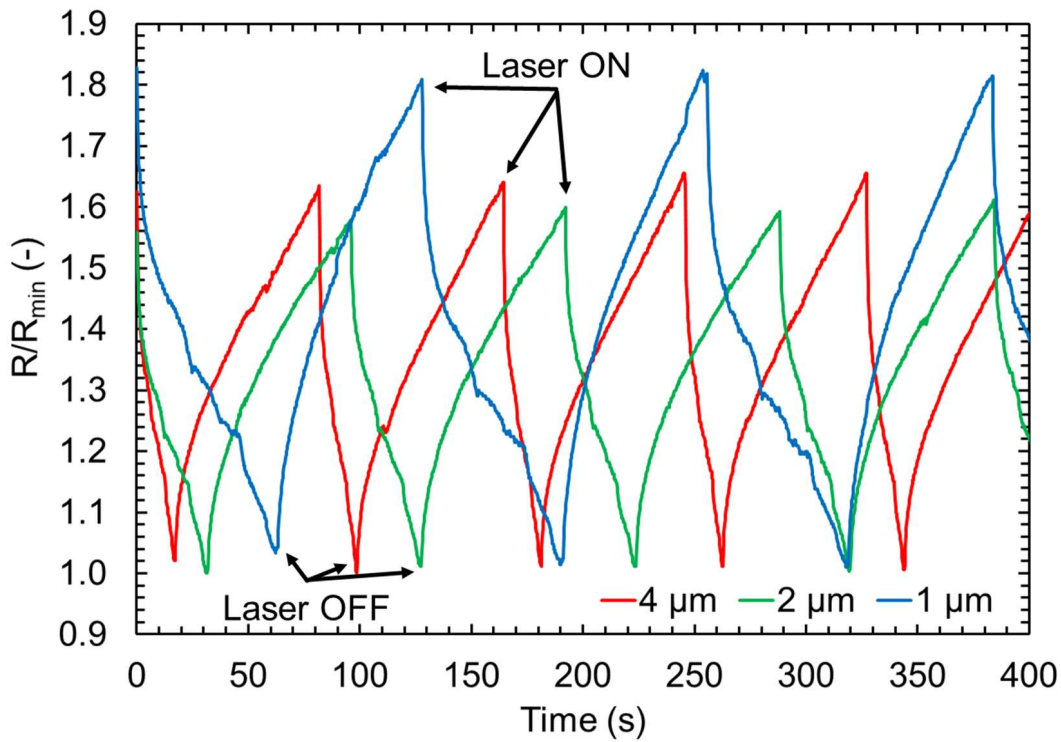

*Figure 6. Variation of resistance (R/R$_{min}$) of ZnO thin film as a function of time when alternatively irradiating with the UV picosecond laser. The samples were maintained in ambient air.*

The first notable outcome from these measurements is the full reversibility of the process, regardless of the hatching distance. For all evaluated hatching distances, similar minimum resistance values ($R_{min}$) in the range of 120 - 160 kΩ were achieved after each cycle of laser annealing. Furthermore, the recovery time required to reach the lowest resistance value can be reduced from 64 seconds to 16 seconds by increasing the hatching distance from 1 µm to 4 µm.

Secondly, it was observed that after processing with hatching distances of 2 µm and 4 µm, the resistance increased up to 1.6 times the $R_{min}$. However, with a hatching distance of 1 µm, the resistance increased to 1.8 times $R_{min}$. This indicates that the resistance increase rate is faster at lower hatching values. This phenomenon can be attributed to the longer processing time associated with smaller hatching distances, during which the thin film may enter in a more reactive state. Consequently, the absorption of oxygen species is accelerated, leading to a more rapid increase in the film's resistance [5].

**4.3.- ZnO thin films for sensible oxygen detection**
The demonstration of the reversibility and recovery of the initial electrical properties opens the door to various applications for these films, particularly in the development of transparent gas sensors. In such applications, the ability to regenerate and reverse processes provides a significant competitive advantage and is a crucial requirement for practical use. For example, a simple transparent oxygen sensor can be fabricated and regenerated by depositing a thin ZnO layer followed by rapid laser irradiation.

The results presented suggest that the time evolution of the electrical properties in laser-treated ZnO films may be intrinsically linked to the oxygen concentration in the surrounding atmosphere [5,8]. Therefore, it is worth exploring the possibility of using these ZnO films to detect oxygen levels.

**Figure 7 (a)** illustrates how the film resistance evolves as a function of the gas concentration, controlled by pressure, within a chamber initially at 30 mbar and subsequently filled with different gases until reaching a pressure of 900 mbar. In this figure, resistance values are normalized relative to the minimum resistance achieved after laser treatment ($R_{min}$). Notably, when argon or nitrogen is introduced into the vacuum chamber, changes in pressure do not affect the rate of electrical conductivity loss, regardless of whether the pressure is increased or decreased. The resistance rates for argon and nitrogen are 0.45 kΩ/s and 0.49 kΩ/s, respectively.

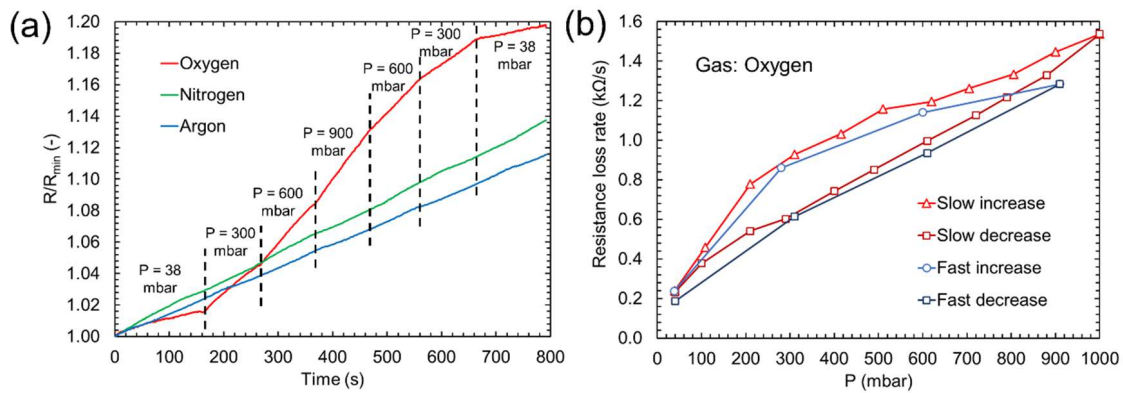

*Figure 7. (a) Time variation of resistance of a ZnO thin film divided by the minimum resistance (R/R$_{min}$). Time zero corresponds to the end of the laser treatment (30 mbar), and once finished, different gases were introduced into the chamber. The gas content is controlled by adjusting the pressure of the studied gas. (b) Resistance loss rate depending on the oxygen concentration inside the chamber.*

However, when oxygen is introduced, there are evident changes in the rate of electrical conductivity variations. The greater the amount of oxygen introduced into the chamber, the higher the rate of the resistance increase. Additionally, it can be observed that when oxygen is removed from the chamber, the rate of loss decreases again, depending exclusively on the amount of oxygen in contact with the ZnO within the controlled atmosphere chamber.

These results suggest that the rate of loss of the electrical resistance is closely related to the oxygen content, exhibiting some differences for increasing or reducing the amount of oxygen (**figure 7 (b)**). Different experiments were conducted to quantify this increase in the electrical resistance over time for varying oxygen concentrations. The rate of resistance increase was measured in an ascending process, gradually increasing the partial pressure of oxygen introduced from vacuum conditions, and in a descending process, progressively reducing the oxygen partial pressure from atmospheric conditions.

The rate of pressure increase/decrease has minimal impact on the resistance rate loss of the thin film, since the same values for different increments of pressure were measured (around 100 mbar for slow increments and 300 mbar for the faster ones). Under high-pressure conditions, variations in this rate become more noticeable. Specifically, the ZnO films exhibit higher resistivity when the pressure increment is applied more gradually. This observation is consistent with the expected behaviour, as a slower increase imposes fewer kinetic limitations on the process, allowing for more controlled and uniform structural modifications within the film.

However, some hysteresis and differences between the pressure increase and decrease phases were observed. These differences are probably attributed to residual oxygen traces inside the vacuum chamber. Such traces are more likely to appear when gas is introduced gradually, possibly due to leaks in the seals or in the chamber, leading to higher resistance rate loss values. Conversely, when the pressure is progressively reduced by extracting gas from the chamber, it is also likely that any residual oxygen is being removed, reducing the resistance rate loss.

Further experiments must be conducted to confirm this hypothesis and verify whether it is indeed possible to reduce hysteresis, which would allow for the precise determination of oxygen present in the described controlled atmosphere environment by monitoring the rate of electrical conductivity loss in the ZnO thin films subjected to the present study. In this way, a transparent oxygen detection sensor based on laser-processed ZnO thin films could be developed. The sensor can be regenerated for reuse by repeating the laser treatment, even with high-speed processes on the order of 16 seconds.

### 4.4.- Influence of the treatment atmosphere on laser processing of ZnO thin films

Based on previous studies, it is clear that the oxygen content in the atmosphere where ZnO thin films are measured significantly influences the temporal evolution of their physical properties. However, in all prior research, the laser processing step was conducted under vacuum conditions (low oxygen content). Therefore, it is essential to assess the influence of the atmospheric environment during laser processing to determine its impact on the final properties of laser-annealed ZnO thin films.

**Figure 8** presents the evolution of electrical resistance over time after processing the thin films in different atmospheres, including air and argon at atmospheric pressure, and at 30 mbar in vacuum conditions. Following the laser processing, the samples were maintained in the selected atmosphere for 10 minutes.

Subsequently, all measurements shown in **figure 8** were conducted under air conditions at atmospheric pressure for all cases studied.

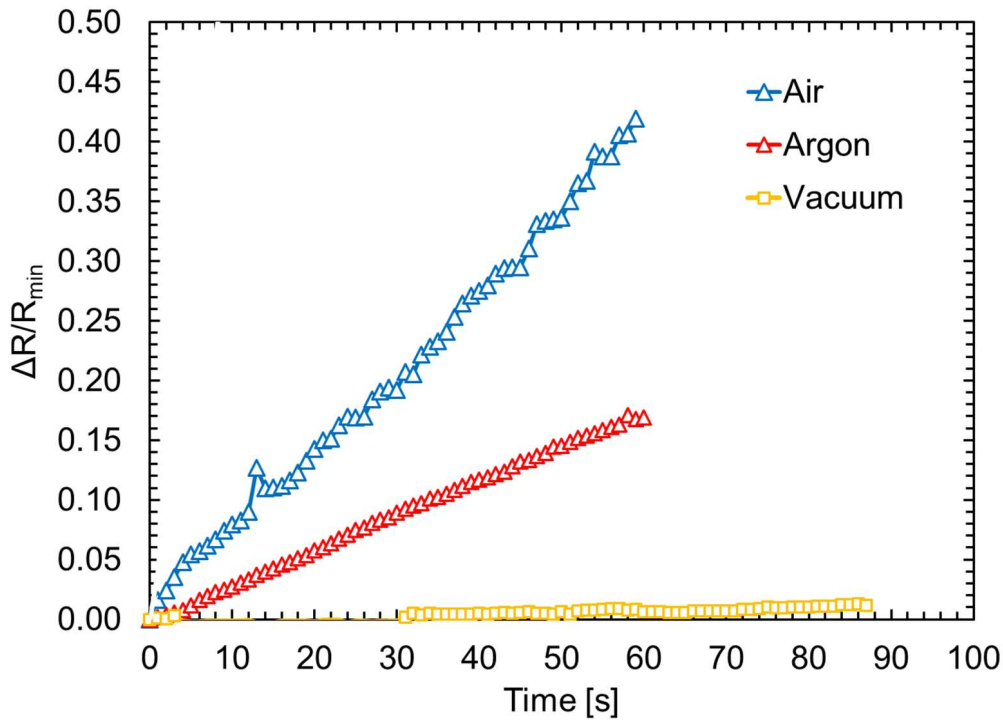

*Figure 8. Evolution with time of the electrical resistance of ZnO thin films after UV laser annealing ($E_p$ = 0.3 μJ, $\delta$ = 1 μm) performed in different atmospheres, including air and argon at atmospheric pressure, and at 38 mbar in vacuum conditions.*

On the one hand, regardless of the treatment atmosphere, the electrical resistance is observed to increase in an approximate linear fashion as a function of time, in accordance with previous experiments. Moreover, these measurements reveal that, when the laser treatments are performed in vacuum, the variation rate is much lower than when they are performed in air or in argon: 1.74, 0.73 and 0.02 kΩ/s are, respectively, the values obtained when the samples were processed in air, argon and vacuum.

Processing in air atmosphere has proven that reactive gases such as $NH_3$, $H_2$, or $CH_4$ interact with the ZnO surface, causing changes in its resistance [8]. The presence of nitrogen and oxygen in the processing atmosphere develops surface dynamics that accelerate the rate of electrical conductivity loss. When processing in argon, a noble gas, the increase in resistance is mitigated compared to processing in air. However, the desorbed oxygen from the film is not evacuated, facilitating reabsorption over time and leading to an intermediate increase rate in resistance. In contrast, processing the sample under dynamic vacuum conditions effectively evacuates the desorbed oxygen and favours preservation of the oxygen vacancies generated during the laser treatment. These phenomena appear to stabilize the surface reactivity, significantly minimizing the temporal increase in resistance when the film is re-exposed to air.

These results open the possibility to introduce a last step to preserve the favourable laser induced properties. This step would involve the *in situ* deposition of a protective coating, such as $Al_2O_3$ [5], immediately after laser processing, and under vacuum conditions. This would ensure that the laser-irradiated ZnO films would not be exposed to oxygen, thereby maintaining the low resistivity values induced by the laser process. Although this approach is not currently feasible with the equipment and conditions available in our laboratory, the findings highlight its potential for producing ZnO-based thin films for applications requiring conductive coatings.

These results illustrate the significance of the treatment and measurement atmosphere and the laser processing parameters in ZnO thin films, as they play a crucial role in the reorganization of oxygen vacancies and other chemical changes in this material.

**5.- Conclusions**

The work reported here demonstrates that laser annealing can tune and modify the resistivity of 90-nm thick ZnO films deposited by SALD. The best electrical performance achieved on these films, after appropriate laser processing conditions yield an electrical resistivity value of $(9 \pm 2) \cdot 10^{-2}$ Ω cm. A wide range of medium laser pulse energies and overlap help maintain the excellent electrical properties of the thin films. If the energy exceeds a certain pulse energy threshold, a progressive loss of the induced electrical properties is observed. This is caused by the deterioration of the thin film integrity, with the initial formation of connected ZnO domains and the eventual loss of continuity between them, upon further energy per pulse increments.

Additionally, an analysis of the applied laser process and the evolution of the electrical resistance of the films as a function of time was performed. A resistance model that reproduces the laser processing experiments and the observed changes on the resistivity over time of the ZnO thin films is herein proposed. After the improvement of the electrical characteristics of the coatings by UV laser irradiation, the electrical resistance starts increasing in a two-stage process: within the first seconds a rapid increase appears to end in a linear trend until an insulating behaviour is reached.

However, the laser annealing has proven to recover the low initial resistivity values induced by the UV laser radiation. Additionally, the recovery time of a 1 $cm^2$ sample can be adjusted and reduced to 16 seconds by increasing the hatching distance to 4 µm, thus reducing the overlap between two laser lines.

Subsequently, the rate of loss of the electrical resistance is observed to be highly dependent on the oxygen concentration in the surrounding atmosphere. The greater the amount of oxygen introduced into the chamber, the higher the rate of the resistance increase. Some hysteresis and differences between the pressure increase and decrease phases were observed and they were attributed to residual oxygen traces inside the vacuum chamber. Ultimately, the laser annealed ZnO thin films were successfully tested as transparent, room-temperature oxygen sensors with a detection limit of 30 mbar constrained by our measurement set-up.

Finally, it has been observed that the treatment atmosphere also affects the rate of ZnO's electrical resistance decay after the treatment. A vacuum atmosphere has been shown to reduce the rate of resistance loss more effectively than argon or air atmospheres. Our findings suggest a promising method to preserve laser-induced properties through the *in situ* deposition of a protective coating under vacuum conditions. This approach could enable the production of ZnO-based conductive thin films.

**Acknowledgements**

The authors gratefully acknowledge the financial support from EU project SPRINT (H2020-FETOPEN 801464), Spanish MCIN/AEI/10.13039/501100011033 (project PID2020-113034RB-I00) and from Gobierno de Aragón (research group T54_23R). Authors also would like to acknowledge the use of Servicio General de Apoyo a la Investigación-SAI and Laboratorio de Microscopías Avanzadas at University of Zaragoza. A. Frechilla and J. Frechilla acknowledge support of the Gobierno de Aragón through their predoctoral contract.

# Supplementary information

A. Frechilla[1,2], J. Frechilla[1], L. A. Angurel[1], G. F. de La Fuente[1] and D. Muñoz-Rojas[2]

[1] Instituto de Nanociencia y Materiales de Aragón, CSIC-Universidad de Zaragoza, María de Luna 3, E-50018 Zaragoza, Spain;
[2] Univ. Grenoble Alpes, CNRS, Grenoble INP, LMGP, F-38000 Grenoble, France;

**Model development**

The calculations for the proposed model are detailed in this section. The proposed model consists of two resistances in parallel: one corresponding to the untreated region ($R_{\text{untreated}}$), and the other to the laser-treated region ($R_{treated}$), as described in **equation A1**.

$$\frac{1}{R} = \frac{1}{R_{\text{untreated}}} + \frac{1}{R_{treated}} \qquad \text{Eq A1}$$

The resistivity ($\rho$) of the untreated region is considered to be the initial resistivity, $\rho_0$, of the ZnO film. In the resistivity calculation of the laser-irradiated region, two effects must be considered: first, the resistivity decreases when the film is irradiated with the laser, modelled by the parameter $\beta_0$, and second, it increases over time due to exposure to the environment. In a first approximation, this increase in resistivity is assumed to follow a radical dependence, determined by an $\alpha$ parameter. Therefore, and with this assumption, the resistivity of the different regions of the ZnO film is defined by **equations A2** and **A3**:

- Untreated region:

$$\rho = \rho_0 \qquad \text{Eq A2}$$

- Laser-treated region:

$$\rho_1(t) = \beta(t)\rho_0 = \beta_0\left(1 + \alpha\sqrt{t}\right)\rho_0 \ \text{with } \beta_0 < 1;\ \alpha > 0 \qquad \text{Eq A3}$$

With these assumptions, the resistance of the untreated region can be calculated using the following relation presented in **equation A4**.

$$\frac{1}{R_{untreated}} = \frac{(W - v_{trans} \cdot t) \cdot e}{\rho_0 \cdot L} \qquad \text{Eq A4}$$

In this equation, $W$ is the total width of the sample, $L$ is the distance between the measurement contacts, $e$ is the thickness of the film, and $V_{trans}$ is the transverse speed of the laser beam.

For the calculation of the resistance of the laser-treated region, it is divided into small regions of width $dx$. Each of these regions was processed at a different instant, so it is necessary to consider the time during which resistance has been increasing in each $dx$ region. The variable $t^*$ has been introduced to consider the period of time between the instant when the laser irradiated a particular region and the considered measurement time, $t$ (**equation A5**).

$$t^* = t - \frac{x}{v_{trans}} \qquad \text{Eq A5}$$

Therefore, the differential of the inverse resistance of the treated region can be expressed following **equation A6**.

$$\frac{1}{R_{treated}} = \int_0^{X'} \frac{e}{\beta_0 \rho_0 L} \frac{dx}{\left(1 + \alpha \sqrt{t - \frac{x}{v_{trans}}}\right)} \qquad \text{Eq A6}$$

In this case, the proposed variable substitution is:

$$t - \frac{x}{v_{trans}} = y^2 \qquad \text{Eq A7}$$

$$2 \cdot y \cdot dy = -\frac{dx}{v_{trans}} \qquad \text{Eq A8}$$

For this variable change, we can calculate the new limits of the definite integral:

$$\begin{cases} \text{When } x = 0 \rightarrow y_0 = \sqrt{t} \\ \text{When } x = X' \rightarrow y_1 = t - \frac{X'}{V_{transversal}} = 0 \end{cases} \qquad \text{Eq A9}$$

Thus, the defined integral with the variable change from **equation A10** can be rewritten as follows.

$$\frac{1}{R_{treated}} = \frac{-2 \cdot v_{trans} \, e}{\beta_0 \rho_0 L} \int_{y_0}^{y_1} \frac{y \cdot dy}{1 + \alpha \cdot y} \qquad \text{Eq A10}$$

Solving the integral, **equation A11** is obtained.

$$\frac{1}{R_{treated}} = \frac{-2 V_{transversal} \, e}{\beta_0 \rho_0 L} \left(\frac{1}{\alpha} y - \frac{1}{\alpha^2} ln(1 + \alpha y)\right)_{\sqrt{t}}^{0} \qquad \text{Eq A11}$$

Therefore, the inverse of the resistance in the treated region, based on this new approximation, is given by the following expression.

$$\frac{1}{R_{treated}} = \frac{-2 \cdot v_{trans} \, e}{\beta_0 \rho_0 L} \left(\frac{1}{\alpha} \sqrt{t} - \frac{1}{\alpha^2} ln(1 + \alpha \sqrt{t})\right) \qquad \text{Eq A12}$$

Finally, substituting the expressions obtained in **equation A4** and **equation A12** into **equation A1** results in a new expression for the inverse of the total resistance:

$$\frac{1}{R} = \frac{(W - v_{trans} \cdot t) \cdot e}{\rho_0 \cdot L} + \frac{-2 \cdot v_{trans} e}{\beta_0 \rho_0 L \alpha} \left(\sqrt{t} - \frac{1}{\alpha} \ln(1 + \alpha\sqrt{t})\right) \qquad \text{Eq A13}$$

By factoring out the constant values, we obtain **equation A14**.

$$\frac{1}{R} = \frac{1}{R_{initial}} \left[1 - \frac{v_{trans}}{W} \cdot t + \frac{2 \cdot v_{trans}}{\beta_0 W \alpha} \left(\sqrt{t} - \frac{1}{\alpha} \ln(1 + \alpha\sqrt{t})\right)\right] \qquad \text{Eq A14}$$

Thus, the final model based on this second approximation is shown in **equation A15**.

$$\frac{R}{R_{initial}} = \frac{1}{1 - \frac{v_{trans}}{W} \cdot t + \frac{2 \cdot v_{trans}}{\beta_0 W \alpha} \cdot \left(\sqrt{t} - \frac{1}{\alpha} \ln(1 + \alpha\sqrt{t})\right)} \qquad \text{Eq A15}$$